\documentclass[a4paper,11pt]{article}
\pdfoutput=1 

\usepackage{jcappub} 

\usepackage[T1]{fontenc} 

\title{Robust Evidence for Dynamical Dark Energy from DESI Galaxy–CMB Lensing Cross-Correlation and Geometric Probes}

\author[a,1]{Miguel A. Sabogal,\note{Corresponding author.}}
\author[a,b]{Rafael C. Nunes}

\affiliation[a]{Instituto de F\'{i}sica, Universidade Federal do Rio Grande do Sul, 91501-970 Porto Alegre RS, Brazil}
\affiliation[b]{Divisão de Astrofísica, Instituto Nacional de Pesquisas Espaciais, Avenida dos Astronautas 1758, São José dos Campos, 12227-010, São Paulo, Brazil}

\emailAdd{miguel.sabogal@ufrgs.br}
\emailAdd{rafadcnunes@gmail.com}

\abstract{Recent analyses joining data from the Cosmic Microwave Background (CMB), Baryon Acoustic Oscillations (BAO), and Type Ia Supernovae (SNIa) have provided strong evidence in favor of dynamical dark energy (DDE) over a simple cosmological constant. Motivated by these findings, we present new observational constraints on DDE based on the cross-correlation between DESI Luminous Red Galaxies (LRG) samples and CMB lensing (\( \mathrm{CMB}_{\kappa} \times \mathrm{LRG} \)), which effectively probes the impact of cosmological parameters on the growth of structure at the perturbative level. We demonstrate that, when combined with geometric measurements such as BAO and SNIa, this cross-correlation yields compelling statistical evidence for DDE exceeding \(4\sigma\), including within simpler parametrizations such as the \( w \)CDM model. Remarkably, this evidence is independent of constraints from primary Planck CMB anisotropies data. These results highlight the robustness and potential of Galaxy–CMB lensing cross-correlation as a powerful observational probe of the dark sector, particularly when used in conjunction with geometric observables.}

\begin{document}
\maketitle
\flushbottom

\section{Introduction}
\label{sec:introduction}

Dark energy (DE) continues to be one of the most profound mysteries in modern astrophysics and cosmology. It is thought to represent an exotic form of energy with negative pressure, responsible for the universe's present accelerated expansion
\cite{DE_obs_1, DE_obs_2}. Describing DE through a cosmological constant ($\Lambda$) offers the simplest and widely accepted explanation for the observed expansion and history of the universe over the last twenty years. A vast amount of observational data--including supernovae, the Cosmic Microwave Background (CMB), Large Scale Structure (LSS) studies, and gravitational lensing (GL)--provides compelling evidence for the presence and influence of DE \cite{Planck:2018vyg,ACT:2020frw,DES:2021bvc,Asgari_2021}. The ability of the $\Lambda$ term to successfully account for these findings has solidified the $\Lambda$CDM model as the dominant paradigm in cosmological research.

In recent years, an increasing number of statistically robust discrepancies and anomalies have emerged from the comparison of diverse cosmological observations. The most important among these is the well-known Hubble tension, an exceeding $5\sigma$ mismatch between the determined value of the Hubble constant, $H_0$, from measurements of the CMB within the standard $\Lambda$CDM framework and the value obtained through local distance ladder techniques employed by the SH0ES team~\cite{Riess:2021jrx}. In addition to the $H_0$ discrepancy, another major issue involves the parameter $S_8$, defined as $S_8 = \sigma_8 \sqrt{\Omega_m / 0.3}$, which combines information about the amplitude of matter clustering ($\sigma_8$) and the matter density parameter $\Omega_m$. Initial analyses of weak lensing data pointed to a tension at the $3\sigma$ level or higher~\cite{KiDS:2020suj,DES:2017qwj}. However, more up-to-date results from the KiDS collaboration have shown consistency with $\Lambda$CDM predictions~\cite{Wright:2025xka,Stolzner:2025htz}. Likewise, measurements of $S_8$ from ACT-CMB observations align well with Planck data, reinforcing the standard cosmological picture~\cite{ACT:2025fju}. Additional surveys have reported findings that support the $\Lambda$CDM scenario in terms of $S_8$ constraints~\cite{Sailer:2024coh,Chen:2024vvk,Anbajagane:2025hlf,Garcia-Garcia:2024gzy,SPT-3G:2025bzu,Sugiyama:2023fzm}. The continued presence of these inconsistencies across multiple independent observations prompts critical reflection on the adequacy of the $\Lambda$CDM framework. This may point to the need for theoretical extensions, including improved modeling of the matter power spectrum, alternative descriptions of DE, modifications to gravitational theory, or the inclusion of possible interactions within the dark sector (see~\cite{CosmoVerse:2025txj} for a comprehensive review).

At the intersection of key cosmic processes, particularly within the context of LSS observations, lies the matter power spectrum. This fundamental tool in cosmology is essential for decoding the relationship between cosmological parameters and the formation of structure in the universe. The power spectrum provides a unique window into both the early and late-time universe, capturing the imprints of initial perturbations, the gravitational influence of dark matter (DM), and the effects of DE. Equally crucial is the role of the non-linear matter power spectrum, which has become indispensable in modern LSS surveys. It offers critical insights into the evolution and complexity of cosmic structures at small and intermediate scales. In this way, the non-linear regime of the power spectrum acts as a vital interface between theoretical models and observational data in contemporary cosmological analyses. Furthermore, the non-linear matter power spectrum serves as a foundational theoretical quantity for interpreting observational measurements, particularly those derived from cross-correlations between different cosmic tracers, our central focus of the present work.

Motivated by recent studies that suggest potential signatures of a dynamical DE component \cite{Scherer:2025esj,Giare:2025pzu,Giare:2024ocw,Giare:2024gpk,Liu:2025mub,Wolf:2025jlc,Ling:2025lmw,Fikri:2024klc,Anchordoqui:2025epz,Dinda:2024}, this work aims to investigate up-to-date observational datasets that are independent of primary CMB anisotropy measurements. Our analysis will focus on combining geometric probes, such as Baryon Acoustic Oscillations (BAO) and Type Ia Supernovae (SNIa), with the cross-correlation between DESI Luminous Red Galaxies (LRG) and CMB lensing maps from Planck PR4 and ACT DR6 \cite{Sailer:2024coh}. This cross-correlation is particularly sensitive to the growth of matter perturbations and can provide valuable insight into the role and behavior of dark components (DM and DE).

Probing the redshift evolution of LSS through the cross-correlation of galaxies with
CMB lensing maps has been of great observational interest in recent years \cite{Kitanidis:2020xno,Hang:2020gwn,White:2021yvw,Krolewski:2021yqy,Qu:2024sfu,Marques:2019aug,Bleem:2012gm,ACT:2023oei,Sailer:2025rks,ACT:2023ipp,Kim:2024dmg,Garcia-Garcia:2021unp}, with wide application for studies in the most diverse perspectives, including tests of general relativity \cite{ACT:2024npz,Wenzl:2024sug}, neutrinos \cite{Tanseri:2022zfe,Giusarma:2018jei}, among other aplications \cite{Bermejo-Climent:2024bcb,Piccirilli:2024xgo,Alonso:2023guh,Nakoneczny:2023nlt}. As shown in the following sections, we find that (\( \mathrm{CMB}_{\kappa} \times \mathrm{LRG} \)) data, when combined with the latest BAO measurements from DESI-DR2 and various Type Ia Supernova samples, provides compelling evidence in favor of a dynamical DE component. Furthermore, our results highlight that the consistency of the galaxy–CMB lensing cross-correlation, particularly when jointly analyzed with complementary cosmological probes, constitutes a powerful observational tool for probing the nature of the dark sector.

The structure of this paper is as follows. In Section~\ref{modeling}, we briefly review how theoretical observables are computed for use in galaxy–CMB lensing cross-correlation analyses. In Section~\ref{data}, we describe the datasets employed and our independent likelihood implementation. Section~\ref{results} presents our main results, and finally, in Section~\ref{conclusion}, we summarize our conclusions and discuss future perspectives.

\section{Cross-Correlation Modeling of Galaxy Surveys and CMB Lensing}
\label{modeling}

The convergence field, $\kappa(\hat{n})$, quantifies the lensing effect caused by LSS on the trajectory of photons traveling from distant sources (e.g., CMB), to the observer in a given direction $\hat{n}$ on the sky. Can be computed under the assumption that transverse angular variations of the gravitational potential dominate over radial variations. In this regime, the line-of-sight projection becomes the leading contribution, effectively suppressing radial oscillations and allowing the convergence field to be expressed as \cite{Bartelmann:1999yn},

\begin{equation}
    \kappa(\hat{n}) =  \int_0^{\chi^*} d\chi^{\prime} \, \chi^{\prime} \left(1-\frac{\chi^{\prime}}{\chi^*}\right) \nabla^2 \Psi\left(\chi^{\prime} \hat{n}, \chi^{\prime}\right)\, ,
\end{equation}
where $\chi$ is the comoving distance, $z^{*} \approx 1100$ denotes the distance to redshift to the surface of last scattering, and $\nabla^2 \Psi(\vec{x}, \chi) = \frac{3}{2c^2} \Omega_{m,0}\, H_0^2 \, (1+z) \, \delta_m(\vec{x}, \chi)$ is the Poisson equation. In this framework, one can directly relate the convergence field to the underlying matter density fluctuations,
\begin{equation}
    \kappa(\hat{n})=\int_0^{\chi^*} d \chi^{\prime} \, \tilde{W}^\kappa\left(\chi^{\prime}\right) \, \delta_m\left(\chi^{\prime} \hat{n}, \chi^{\prime}\right) \,.
\end{equation}
Here, we define the CMB lensing projection kernel as, 

\begin{equation}
     \tilde{W}^\kappa (\chi) = \frac{3}{2c^2} \Omega_{m,0} H_0^2 \, (1+z) \, \chi \left( 1-\dfrac{\chi}{\chi^{*}}\right) \, ,
\end{equation}
where $\Omega_{m,0}$ is the present-day matter density parameter, $c$ is the speed of light, and $H_0$ is the Hubble constant. To relate the convergence field to observable tracers of LSS, we consider the galaxy density contrast projected onto the sky, obtained by integrating the three-dimensional galaxy overdensity $\delta_g(\vec{x})$ along the line of sight \cite{Crittenden:1995ak}:

\begin{equation}
\delta_g(\hat{n}) = \int d\chi' \, \tilde{W}^g(\chi') \, \delta_g(\chi' \hat{n}, \chi')\, ,
\end{equation}
where the galaxy window function is defined as:

\begin{equation}
\tilde{W}^g(\chi) = \phi(z) \, \dfrac{dz}{d\chi} = \phi(z) \dfrac{H(z)}{c}\, .
\end{equation}

In the above equations, $\phi(z)$ is the normalized redshift distribution of the galaxy sample, satisfying $\int dz \, \phi(z) = 1$. Assuming a linear bias relation between galaxies and matter, $\delta_g(\vec{x}) = b(z) \, \delta_m(\vec{x})$, the projected LSS density contrast on the sky becomes:

\begin{equation}
\delta_g(\hat{n}) = \int d\chi' \, \tilde{W}^g(\chi') \, b(z') \, \delta_m(\chi' \hat{n}, \chi')\, .
\end{equation}

Having expressed both $\kappa(\hat{n})$ and $\delta_g(\hat{n})$ as integrals over the matter density contrast, one can compute their cross-correlation in harmonic space. Following the standard approach \cite{Bartelmann:1999yn,Lewis:2006fu}, the angular cross-power spectrum, now in terms of redshift, is given by:
\begin{equation}
\label{c_ell_general}
C_{\ell}^{\kappa g} = \frac{2}{\pi} \int_0^{z^*} \dfrac{c \, dz}{H(z)} \, \tilde{W}^{\kappa}(z) \, \tilde{W}^{g}(z) \, b(z) \int dk \, k^2 \, j_{\ell}(k\chi) j_{\ell}(k\chi) P_m(k,z),
\end{equation}
where the inner integral over $k$ accounts for the projection of 3D clustering onto the 2D sky, $P_m(k,z)$ is the matter power spectrum, i.e, $\langle \delta_m(k,z) \delta_m^*(k',z') \rangle = (2\pi)^3 \delta(k-k') \delta(z-z') P_m(k,z)$, and $j_\ell$ are spherical Bessel functions. 

Applying the Limber approximation~\cite{1953ApJ...117..134L,LoVerde:2008re}, the angular cross-spectrum for two cosmological tracers, $X$ and $Y$, can be re-write in the general form:
\begin{equation}
C^{XY}_{\ell} = \int dz \, \frac{H(z)}{c} \, \frac{W^{X}(z) W^{Y}(z)}{\chi^2} P_{XY}\left(k = \dfrac{(\ell + 1/2)}{\chi} ; z\right).
\end{equation}

To clarify our notation, in the context of the angular power spectra $C_\ell^{XY}$, the indices $X$ and $Y$ refer to the projected fields involved: $\kappa$ denotes the CMB lensing convergence and $g$ galaxy over-density. When referring to the three-dimensional power spectra \( P_{XY}(k, z) \), the indices correspond to the physical sources of these fluctuations: \( m \) stands for the matter density field and \( g \) for the galaxy distribution (see, e.g., Eqs. \ref{eq:Pmg} and \ref{eq:Pgg}). Note, that now we redefine the projection kernels for our case of interest, corresponding to galaxy overdensity and CMB lensing convergence, respectively:

\begin{equation}
W^g(z) = \phi(z)\, ,
\end{equation}

\begin{equation}
W^\kappa(z) = \frac{3}{2c} \Omega_{m,0} \frac{H_0^2}{H(z)} (1 + z) \, \chi \left( 1-\dfrac{\chi}{\chi^{*}}\right) .
\end{equation}

Therefore, the expressions for the angular power spectra under the Limber approximation yield the following results:

For the galaxy–CMB lensing cross-correlation:
\begin{equation}
C^{\kappa g }_{\ell} = \int_0^{z^*} dz \, \frac{3 \Omega_{m,0} H_0^2}{2 c^2} \frac{(1 + z)}{\chi(z)} \frac{[\chi^* - \chi(z)]}{\chi^*} \, \phi(z) \, P_{mg}\left(k = \frac{\ell + 1/2}{\chi(z)} ; z\right),
\end{equation}

For the galaxy auto-correlation:
\begin{equation}
C^{gg}_{\ell} = \int dz \, \frac{H(z)}{c} \, \frac{\phi^2(z)}{\chi^2(z)} \, P_{gg}\left(k = \frac{\ell + 1/2}{\chi(z)} ; z\right),
\end{equation}

And for the CMB lensing auto-correlation:
\begin{equation}
C^{\kappa \kappa}_{\ell} = \int_0^{z^*} dz \, \frac{H(z)}{c} \left[ \frac{3 \Omega_{m,0} H_0^2}{2 c H(z)} (1 + z)\frac{[\chi^* - \chi(z)]}{\chi^*} \right]^2 P_{mm}\left(k = \frac{\ell + 1/2}{\chi(z)} ; z\right).
\end{equation}

Another important phenomenon is magnification bias, a well-known effect (see \cite{1980ApJ...242L.135T}) that alters the observed number density of galaxies in flux-limited surveys. Gravitational lensing along the line of sight both stretches the apparent area (reducing galaxy counts) and increases the brightness of background galaxies (bringing fainter galaxies above the detection threshold). These competing effects result in correlations between background galaxies and foreground matter, potentially biasing galaxy–galaxy and galaxy–convergence angular power spectra \cite{Ziour:2008awn}. In practice, magnification bias introduces an additional contribution to the galaxy window function:
\begin{equation}
W^g(z) \to W^g(z) + W^\mu(z),
\end{equation}
where, to first order, the magnification term is given by:
\begin{equation}
W^\mu(z) = (5s_{\mu} - 2) \frac{3}{2c} \Omega_{m0} \frac{H_0^2}{H(z)} (1 + z) \int_z^{z^*} dz' \, \frac{\chi(z) \left[\chi(z') - \chi(z)\right]}{\chi(z')} \phi(z')\, .
\end{equation}

Here, $s_{\mu}$ denotes the slope of the cumulative magnitude distribution and characterizes the sensitivity of the observed number density to flux magnification near the survey’s limiting magnitude (see, e.g., \cite{Krolewski:2019yrv} for its importance). This term contributes additional correlations in the angular power spectra, such that the galaxy auto- and cross-spectra are modified as:
\begin{align}
C^{\kappa g}_{\ell} &\to C^{\kappa g}_{\ell} + C^{\kappa \mu}_{\ell}, \\
C^{gg}_{\ell} &\to C^{gg}_{\ell} + 2C^{g \mu}_{\ell} + C^{\mu \mu}_{\ell}.
\end{align}

In a framework aimed at modeling the galaxy-galaxy and matter-galaxy power spectra, $P_{gg}(k)$ and $P_{mg}(k)$, the angular power spectra $C^{gg}_{\ell}$ and $C^{\kappa g}_{\ell}$ serve as valuable observables for constraining both cosmological parameters and galaxy bias. A widely used and straightforward approach to model these spectra involves the application of the HaloFit fitting formula presented by \citep{Takahashi_2012}, which describes the nonlinear matter power spectrum, $P_{mm}^{\text{HF}}(k)$. Using this, we express the galaxy-galaxy and matter-galaxy power spectra as scale-dependent, bias-modulated versions of the matter power spectrum:

\begin{align}
    P_{gg}(k, z) &= b(z)^{2} \, P_{mm}^{\text{HF}}(k, z), \label{eq:Pgg} \\
    P_{mg}(k, z) &= b(z) \, P_{mm}^{\text{HF}}(k, z), \label{eq:Pmg}
\end{align}
where $b(z)$ represents the linear bias parameter at a given redshift $z$, which encodes the correlation between the galaxy distribution and the underlying matter distribution. In this approach, the galaxy bias is assumed to be scale-independent, a sufficient condition for our case of interest in this present work.

To compute the nonlinear matter power spectrum $P_{mm}^{\text{HF}}(k)$, we use the Boltzmann code \texttt{CLASS} \citep{Blas:2011rf}, which provides the necessary cosmological background for accurate power spectrum calculations. For each redshift distribution $\phi(z_{i})$, which is derived from photometric data (see, \textsection\ref{cross-samples} for details), we adopt a bias evolution model where the bias parameters $b(z_{i})$ are assumed to scale with the growth factor $D(z)$ as $b(z_{i}) \propto D(z)$, where $D(z)$ is the linear growth factor, normalized such that $D(0) = 1$ at the present day. This scaling reflects the fact that the galaxy bias typically evolves with the expansion of the universe and the growth of structure. In practice, we simplify the calculation by approximating the Limber integrals, which are typically used to account for redshift dependencies in angular power spectra, by evaluating $P_{gg}$ and $P_{mg}$ at an effective redshift, {therefore, $D(z_{i})$ is just absorbed into the bias coefficient}. This approximation assumes that the redshift evolution of these spectra is sufficiently small over the integration range, making the effective redshift a good approximation for performing the integrals. The effective redshift, $z_{\mathrm{eff}}$, is computed as \cite{White_2022}

\begin{equation}
z_{\mathrm{eff}} = \frac{\int d \chi z(\chi) [ \tilde{W}^g(z) ]^2 / \chi^2}{\int d \chi [ \tilde{W}^g(z) ]^2 / \chi^2},
\label{effective_z}
\end{equation}
where the integrals are weighted by the galaxy window function $\tilde{W}^g(z)$ and the comoving distance $\chi(z)$. This procedure allows for a more efficient computation of the angular power spectra while retaining the essential physical information related to galaxy clustering and the distribution of dark matter.

In summary, by combining the nonlinear matter power spectrum described by HaloFit with the linear bias model, we can effectively model the galaxy-galaxy and matter-galaxy power spectra, enabling us to obtain cosmological constraints from the angular power spectra of galaxies and their correlations with gravitational lensing. Although we incorporate the nonlinear matter power spectrum in our calculations, as detailed in the following section, our analysis will focus primarily on data within the linear regime.

\section{Datasets and methodology}
\label{data}

To test the theoretical structures explored in this work, we implemented our model in the \texttt{CLASS} Boltzmann solver~\cite{Blas:2011rf} and used the \texttt{MontePython} sampler~\cite{Brinckmann:2018cvx, Audren:2012wb} to perform Monte Carlo analyses via Markov Chains (MCMC). 

In this work, we consider a widely used class of DDE models known as the Chevallier–Polarski–Linder (CPL) parameterization \cite{Chevallier:2000qy,Linder:2002et,Linder:2024rdj}. This parametrization is frequently adopted in the literature due to its simplicity and its ability to capture the potential time evolution of the dark energy equation of state. In the CPL model, the equation of state parameter is expressed as a function of the scale factor \( a \) by $w(a) = w_0 + w_a (1 - a)$, where \( w_0 \) represents the present-day value of the equation of state, and \( w_a \) quantifies its rate of change with time. For convenience, we refer to this class of models throughout the text as the $w_0$$w_a$CDM model. In a second stage of our analysis, we will focus on a simpler subclass of this model by fixing \( w_a = 0 \), which corresponds to a constant equation of state $w(a) = w_0$. This reduced case is known in the literature as the $w$CDM model and serves as a natural baseline for assessing the impact of DE on cosmological observables.

\begin{table}[htpb!]
    \centering
    \renewcommand{\arraystretch}{1.25}
    \begin{tabular}{cc||cc}
\hline\hline
\multicolumn{2}{c||}{Cosmological} & \multicolumn{2}{c}{Nuisance} \\
\hline
$10^2 \Omega_{\rm b} h^2$ & $\mathcal{U}(1,5)$ & $b(z_{i})$ & $\mathcal{U}(0.0,5.0)$ \\
$\Omega_{\rm c} h^2$ & $\mathcal{U}(0.08, 0.16)$ & $s_{\mu}(z_{i})$ & $\mathcal{N}(\text{Table~\ref{bins}}, 0.1)$ \\
$\ln(10^{10} A_{\rm s})$ & $\mathcal{U}(2.0, 4.0)$ & ${\rm SN}^{2\rm D}(z_{i})$ & $\mathcal{N}_{\mathrm{r}}(\text{Table~\ref{bins}}, 0.3)$ \\
$H_0 \ [\mathrm{km/s/Mpc}]$ & $\mathcal{U}(20, 100)$ & & \\
$w_0$ or $w$ & $\mathcal{U}(-3.0, 1.0)$ & & \\
$w_a$ & $\mathcal{U}(-3.0, 2.0)$ & & \\
\hline\hline
\end{tabular}
\caption{Flat priors $\mathcal{U}$ used in our fits for the cosmological and nuisance parameters. For each redshift bin, we center prior means for the number
count slope $s_\mu\left(z_i\right)$, and shot noise $\mathrm{SN}^{2 \mathrm{D}}\left(z_i\right)$ around those listed in Table \ref{bins}, following the methodology stated in~\cite{Sailer:2024coh}. Here, $\mathcal{N}(\mu, \sigma)$ denotes a Gaussian prior with mean $\mu$ and standard deviation $\sigma$, and $\mathcal{N}_r(\mu, \sigma) \equiv$ $\mathcal{N}(\mu, \sigma \mu)$.} 
\label{priors}
\end{table}

The cosmological parameters sampled in this work include: the physical baryon density $\omega_{\rm b} = \Omega_{\rm b} h^2$, the physical dark matter density $\omega_{\rm c} = \Omega_{\rm c} h^2$, the amplitude of the primordial scalar power spectrum $A_{\mathrm{s}}$, and the Hubble constant $H_{0}$. In addition to these standard parameters, we also sample $w_{0}$ and $w_{a}$ within the context of DDE models. The priors for all cosmological parameters are listed in the left section of Table~\ref{priors}. In all our joint analyses, convergence was verified using the Gelman–Rubin diagnostic~\cite{Gelman_1992}, requiring that \( R - 1 \leq 10^{-2} \) for all parameter chains. The statistical analysis was carried out using the \texttt{GetDist} package,\footnote{\url{https://github.com/cmbant/getdist}}, which enabled the extraction of numerical results, including one-dimensional posterior distributions and two-dimensional marginalized probability contours. In the following sections, we describe the likelihood functions and the analysis methodology adopted throughout this work.


\subsection{DESI Galaxy–CMB Lensing Cross-Correlations Samples}
\label{cross-samples}
To incorporated the \texttt{CMB$\kappa$\scalebox{1.2}{$\times$}LRG} data in our work, we followed the methodology outlined in \cite{Sailer:2024coh}, and constructed our likelihood within the \texttt{MontePython} framework, using photometric samples of Luminous Red Galaxies (LRGs) from the DESI Legacy Imaging Survey DR9 \cite{Zhou:2023gji, DESI:2022gle, dey2019overview}, combined with CMB lensing convergence maps reconstructed from data by Planck \cite{Carron:2022eyg} and the Atacama Cosmology Telescope (ACT) \cite{ACT:2023dou}. The measurements consist of mask-deconvolved angular power spectra, including the galaxy–CMB lensing cross-spectra ($C_{L}^{\kappa g}$) and the galaxy auto-spectra ($C_{L}^{gg}$), computed across four redshift bins distributions in the range $0.2 < z < 1.0$, along with their associated covariance matrices, pixelization corrections, and window functions. 


\begin{table}[htbp]
\centering
\renewcommand{\arraystretch}{1.3}
\begin{tabular}{c||ccc||ccc}
\hline\hline
Sample 
& \multicolumn{3}{c||}{LRG Properties} 
& \multicolumn{3}{c}{Multipole Cuts} \\
\cline{2-7}
& $z_{\text{eff}}$ & $s_\mu$ & $10^6 \mathrm{SN}^{2\mathrm{D}}$ 
& $\ell_{\min}^{\kappa g}$ (Planck / ACT) & $\ell_{\min}^{gg}$ & $\ell_{\max}$ \\
\hline
$z_2$ & 0.625 & 1.044 & 2.25 & 20 / 44 & 79 & 178 \\
$z_3$ & 0.785 & 0.974 & 2.05 & 20 / 44 & 79 & 243 \\
$z_4$ & 0.914 & 0.988 & 2.25 & 20 / 44 & 79 & 243 \\
\hline\hline
\end{tabular}
\caption{Summary of LRG sample properties and the adopted multipole cuts for the cross- and auto-correlation measurements, based on the analysis of \cite{Sailer:2024coh}.}
\label{bins}
\end{table}

For our analyses, we restrict the implementation to the three highest redshift bins $z_{i}$ ($i=2,3,4$).  As described in Section~\textsection\ref{modeling}, we evaluate the matter–galaxy and galaxy–galaxy power spectra at the effective redshift of each bin, listed in the left section of Table~\ref{bins}. We also adopt the multipole cuts used in the original analysis: $\ell_{\rm min} = 20$ for Planck, $\ell_{\rm min} = 44$ for ACT, and $\ell_{\rm min} = 79$ for $gg$. The maximum multipoles are set to $\ell_{\rm max} = 178$, $243$, and $243$ for bins $z_2$, $z_3$, and $z_4$, respectively, as shown in the right section of Table \ref{bins}. Following the bandpower conventions of \texttt{NaMaster} \cite{Alonso:2018jzx}, we convolve our theoretical predictions with the corresponding window functions, $C_{L}^{XY}=\sum_{\ell} W_{L \ell}^{XY} C_{\ell}^{X Y}$, and truncate the sum at $\ell_{\rm max} = 300$, since $W_{L \ell}^{XY}$ contributions from $\ell > 300$ are negligible in our linear theory implementation. No significant deviations were observed when extending to higher values. We also account for shot noise and pixelization effects by including the appropriate pixel transfer function\footnote{The combination of these effects can be well approximated by taking $C_{\ell}^{\kappa g} \rightarrow w_{\ell} C_{\ell}^{\kappa g} \text { and } C_{\ell}^{g g} \rightarrow w_{\ell}^2 C_{\ell}^{g g}+\mathrm{SN}^{2 \mathrm{D}}$, where \( w_{\ell} \) is the pixel window function and \( \mathrm{SN}^{2\mathrm{D}} \) represents the shot noise contribution.} for each $C_{\ell}^{XY}$, which mildly suppresses the power at high multipoles.

To extract cosmological information from our measurements, we perform a joint analysis of the galaxy clustering and galaxy-CMB lensing correlations, i.e., we simultaneously fit the galaxy-galaxy auto-spectrum and the galaxy-convergence cross-spectrum using a Gaussian likelihood function:
\begin{equation}
    \mathcal{L}(\mathbf{C} | \boldsymbol{\theta}) \propto \exp \left( 
    -\frac{1}{2} \left[ \hat{\mathbf{C}}(\boldsymbol{\theta}) - \mathbf{C} \right]^\top \mathbf{\Sigma}^{-1} \left[ \hat{\mathbf{C}}(\boldsymbol{\theta}) - \mathbf{C} \right]
    \right),
\end{equation}
where \(\boldsymbol{\theta}\) denotes the set of model parameters, $\mathbf{C}$ and $\hat{\mathbf{C}}(\boldsymbol{\theta})$ are the observed and predicted angular power spectra, respectively. The combined auto- and cross-spectrum measurements take the form $\mathbf{C}_{L} = \left(C_{L}^{g g},C_{L}^{\kappa  g} \right)^\top$, and the covariance matrix \(\boldsymbol{\Sigma}\) encapsulates both the statistical uncertainties and their correlations between different multipole bins $L$. 

By adopting the cosmological priors on $\Omega_c h^2$ and $\ln(10^{10} A_s)$ listed in the right-hand section of Table~\ref{priors}, along with the nuisance parameter priors from the same table, we successfully reproduced the Linear Theory results of~\cite{Sailer:2024coh} (see Figure~\ref{linear_theory}). 

{In Appendix \ref{Appendix}, we demonstrate that alternative modeling choices for the matter power spectrum $P(k,z)$ do not significantly affect our main results. Specifically, we assess the impact of using different prescriptions, such as \texttt{HMcode} \cite{Mead:2016zqy,Mead:2020vgs} and the inclusion of counterterm contributions, and find that, within the data ranges and scales considered in this work, all these approaches yield statistically consistent outcomes. This robustness underscores that our conclusions are not sensitive to the specific nonlinear modeling employed. For consistency and simplicity, we proceed with Halofit as our default prescription throughout the remainder of the analysis.}


\begin{figure*}[htpb!]
    \centering
    \includegraphics[width=0.75\textwidth]{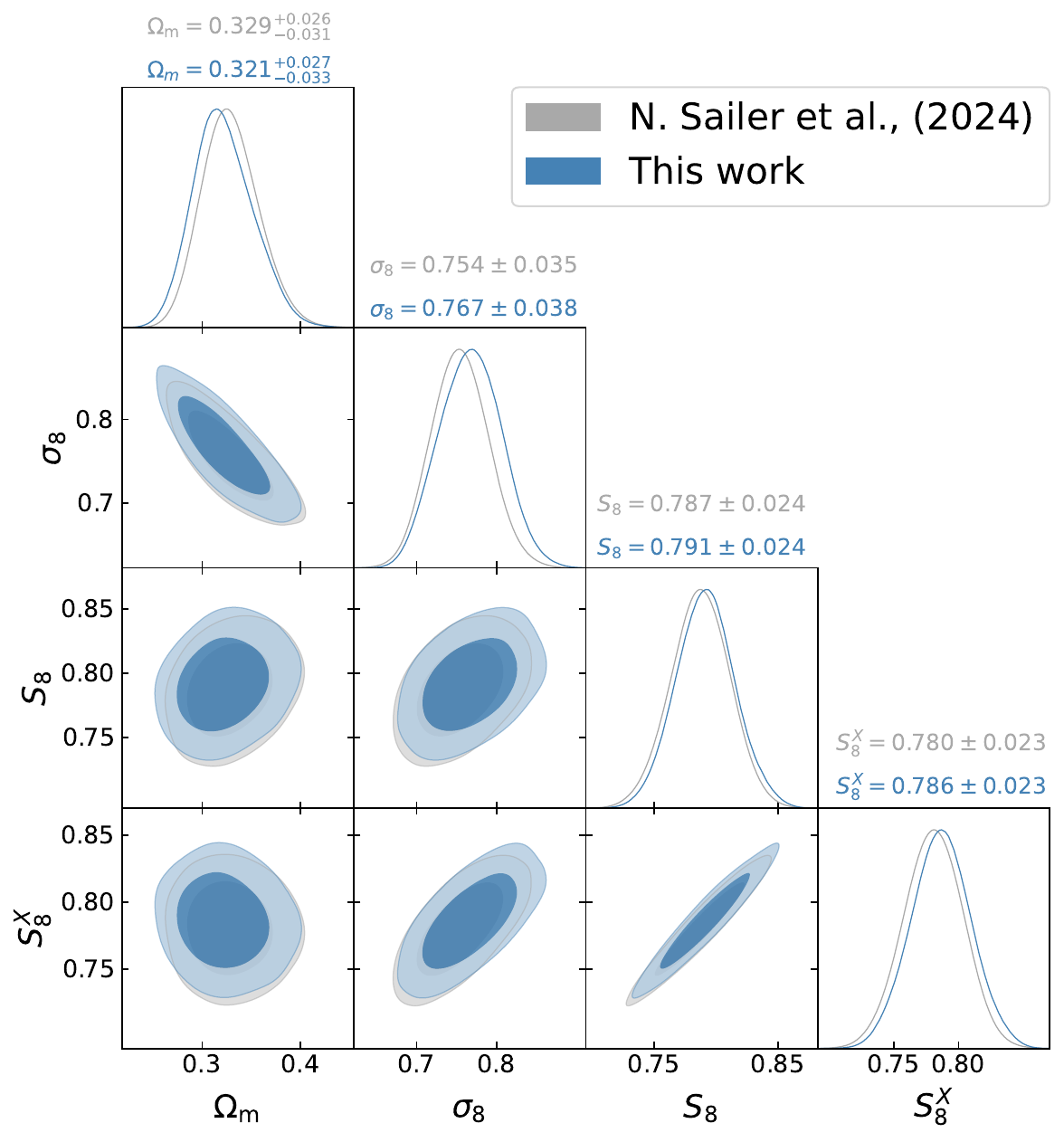}
    \caption{Triangle plot comparing the posterior distributions of cosmological parameters obtained in this work using our independently implemented likelihood using Monetpython (\textcolor[rgb]{0.27, 0.51, 0.71}{blue}) with those reported by \cite{Sailer:2024coh} (gray). The parameters shown include the matter density parameter \(\Omega_{\rm m}\), the amplitude of matter fluctuations \(\sigma_8\), the derived parameters \(S_8\) and $S_8^X = \sigma_8 (\Omega_m/0.3)^{0.4}$. The contours correspond to the 68\% and 95\% confidence levels. The agreement between the two results serves as a consistency check and confirms the accuracy of our likelihood implementation.}
    \label{linear_theory}
\end{figure*}

\subsection{Additional data}

The additional datasets used in our joint analyses are:

\begin{itemize}


\item \textit{Baryon Acoustic Oscillations} (\textbf{DESI-DR2}): We use BAO measurements from the second data release of the DESI survey, which includes data from galaxies, quasars~\cite{DESI:2025zgx}, and Lyman-$\alpha$ tracers~\cite{DESI:2025zpo}. These measurements, detailed in Table IV of Ref.~\cite{DESI:2025zgx}, span the {effective redshift range} $0.295 \leq z \leq 2.330$, divided into nine bins. BAO constraints are presented in terms of the transverse comoving distance $D_{\mathrm{M}}/r_{\mathrm{d}}$, the Hubble horizon $D_{\mathrm{H}}/r_{\mathrm{d}}$, and the angle-averaged distance $D_{\mathrm{V}}/r_{\mathrm{d}}$, all normalized to the comoving sound horizon at the drag epoch, $r_{\mathrm{d}}$. We also account for correlation structures through cross-correlation coefficients: $r_{V,M/H}$ and $r_{M,H}$, which capture the relationships between different BAO measurements. This dataset is referred to as \texttt{DESI-DR2}.

\item \textit{Type Ia Supernovae} (\textbf{SN Ia}): We utilize the following recent samples of SN Ia:

\begin{enumerate}
    \item[(i)] \textbf{PantheonPlus}: The PantheonPlus sample~\cite{Brout:2022vxf} provides distance modulus measurements from 1701 light curves of 1550 distinct SN Ia events, covering the redshift range $0.01 \leq z \leq 2.26$. This dataset is referred to as \texttt{PP}.

    \item[(ii)] \textbf{Union 3.0}: The Union 3.0 compilation consists of 2087 SN Ia in the redshift range $0.001 < z < 2.260$~\cite{Rubin:2023ovl}, with 1363 overlapping with the PantheonPlus sample. This dataset, referred to as \texttt{Union3}, uses Bayesian hierarchical modeling to address systematic uncertainties and errors.

    \item[(iii)] \textbf{DESY5}: As part of the Year 5 data release, the Dark Energy Survey (DES) presents a new sample of 1635 photometrically classified SN Ia with redshifts $0.1 < z < 1.3$~\cite{DES:2024jxu}, along with 194 low-redshift SN Ia (shared with the PantheonPlus sample) in the range $0.025 < z < 0.1$. This dataset is referred to as \texttt{DESY5}.
\end{enumerate}

\end{itemize}

As a final note, we emphasize that throughout our analysis we adopt state-of-the-art assumptions for Big Bang Nucleosynthesis (BBN), sensitive to the constraints on the physical baryon density $\omega_{\rm b} = \Omega_{\rm b} h^2$. Specifically, the \texttt{BBN} likelihood that incorporates the most precise available measurements of primordial light element abundances: the helium mass fraction \( Y_P \), as determined in~\cite{Aver_2015}, and the deuterium-to-hydrogen ratio \( y_{\rm DP} = 10^5\, n_D / n_H \), from~\cite{Cooke_2018}. 

\section{Results}
\label{results}

In the following, we present and discuss our main results. We begin by outlining the criteria adopted to ensure a robust comparison between the theoretical models and the statistical analyses. Subsequently, we present the statistical results obtained from our analysis.

\subsection{Bayesian Model Comparison}

To better evaluate the consistency between models and their compatibility with the observational datasets considered, we perform a statistical comparison between the DDE model and the standard $\Lambda$CDM scenario. For this purpose, we adopt two complementary model selection criteria: the Akaike Information Criterion (AIC) and the Bayesian Evidence (BE).

The AIC \cite{AIC} is usually defined as:
\begin{equation}
\mathrm{AIC} \equiv -2 \ln \mathcal{L}_{\text{max}} + 2N,
\label{AIC}
\end{equation}
where $\mathcal{L}_{\text{max}}$ denotes the maximum likelihood achievable by the model, and $N$ is the total number of free parameters. A lower AIC value indicates a better model, balancing goodness of fit and model complexity. Specifically, the AIC includes a penalty for models with more parameters, discouraging overfitting.

In addition to the AIC, we also consider the Bayesian Evidence, which provides a more rigorous assessment by computing the Bayes factor, quantifying the support for one model relative to another. Given a dataset $\mathbf{x}$ and two competing models, $\mathcal{M}_i$ and $\mathcal{M}_j$, with parameter sets $\boldsymbol{\theta}_i$ and $\boldsymbol{\theta}_j$, respectively, the Bayes factor $\mathcal{B}_{ij}$ (assuming equal prior probabilities for the models) is defined as:

\begin{equation}
\mathcal{B}_{ij} = \frac{p(\mathcal{M}_i|\mathbf{x})}{p(\mathcal{M}_j|\mathbf{x})} = \frac{\displaystyle \int d \boldsymbol{\theta}_i \, \pi\left(\boldsymbol{\theta}_i | \mathcal{M}_i\right) \mathcal{L}\left(\mathbf{x} | \boldsymbol{\theta}_i, \mathcal{M}_i\right)}{\displaystyle\int d \boldsymbol{\theta}_j \, \pi\left(\boldsymbol{\theta}_j | \mathcal{M}_j\right) \mathcal{L}\left(\mathbf{x} | \boldsymbol{\theta}_j, \mathcal{M}_j\right)},
\label{BayesFactor}
\end{equation}

where $p(\mathcal{M}_i|\mathbf{x})$ is the Bayesian evidence for model $\mathcal{M}_i$, $\pi(\boldsymbol{\theta}_i|\mathcal{M}_i)$ denotes the prior distribution, and $\mathcal{L}(\mathbf{x}|\boldsymbol{\theta}_i, \mathcal{M}_i)$ is the likelihood of the data given the parameters. A Bayes factor $\mathcal{B}_{ij} > 1$ favors model $\mathcal{M}_i$ (e.g., $\Lambda$CDM) over model $\mathcal{M}_j$ (e.g., DDE), even if the latter has a better fit, due to penalization for model complexity and prior volume.

To interpret the strength of evidence, we adopt the Jeffreys–Kass–Raftery scale~\cite{Kass:1995loi}:
\begin{itemize}
    \item $0 \leq |\ln \mathcal{B}_{ij}| < 1$: Weak evidence.
    \item $1 \leq |\ln \mathcal{B}_{ij}| < 3$: Positive evidence.
    \item $3 \leq |\ln \mathcal{B}_{ij}| < 5$: Strong evidence.
    \item $|\ln \mathcal{B}_{ij}| \geq 5$: Very strong evidence.
\end{itemize}

To summarize, a negative value of $\ln \mathcal{B}_{ij}$ suggests a preference for the DDE model over $\Lambda$CDM, whereas a positive value favors $\Lambda$CDM. For computing Bayes factors, we use the publicly available \texttt{MCEvidence} package~\cite{Heavens:2017hkr,Heavens:2017afc}.\footnote{Available at \url{https://github.com/yabebalFantaye/MCEvidence}.}

\subsection{$w_{0}w_{a}$CDM}

Tables~\ref{table_w0wa} and~\ref{table_DEE} summarize the constraints at 68\% confidence level (CL) for the baseline parameters in the two models considered in this work, namely the \( w_0w_{\rm a} \)CDM and \( w \)CDM models, respectively. The results are presented for all joint analyses performed, as indicated in the table headers. In our tests, we find that the combination \( \mathrm{CMB}_{\kappa} \times \mathrm{LRG} \) alone does not have sufficient constraining power to simultaneously determine all free parameters in the \( w_0w_{\rm a} \)CDM model. Therefore, we only present results for this case based on the joint analysis.

\begin{table*}[htpb!]
\centering
\caption{Marginalized constraints—mean values with 68\% confidence levels—on the free and selected derived parameters of the $w_0w_{\rm a}$CDM model are presented for the $\mathrm{CMB}_{\kappa} \times \mathrm{LRG}$ dataset and its combinations with BAO-DESI-DR2, PantheonPlus (PP), Union3, and DES Y5. The last rows report the differences in minimum chi-square values, $\Delta \chi^2_{\rm min} = \chi^2_{\rm min}(w_0w_{\rm a}\mathrm{CDM}) - \chi^2_{\rm min}(\Lambda\mathrm{CDM})$, the Akaike Information Criterion difference, $\Delta \mathrm{AIC} = \mathrm{AIC}_{w_0w_{\rm a}\mathrm{CDM}} - \mathrm{AIC}_{\Lambda\mathrm{CDM}}$, and the Bayes factors $\ln B_{ij}$, as defined in Eq.~(\ref{BayesFactor}). Negative values of $\Delta \chi^2_{\rm min}$ and $\Delta \mathrm{AIC}$ indicate a better fit of the $w_0w_{\rm a}$CDM model compared to the $\Lambda$CDM model, while a negative $\ln B_{ij}$ implies a preference for the $w_0w_{\rm a}$CDM model $\Lambda$CDM.}
\renewcommand{\arraystretch}{1.55}
\resizebox{\textwidth}{!}{
\begin{tabular}{l||c|ccc} 
\hline
\textbf{Parameter} & \textbf{CMB$\kappa$\scalebox{1.2}{$\times$}LRG + DESI-DR2} & \textbf{+ PP} & \textbf{+ Union3} & \textbf{+ DESY5} \\ 
\hline \hline
$10^{2} \Omega_{\rm b} h^2$  &  $2.271\pm 0.038$ & $2.270^{+0.033}_{-0.040}$ & $2.272\pm 0.037$ & $2.271\pm 0.038$  \\

$\Omega_{\rm c} h^2$  &  $0.1309 \pm 0.0057$ & $0.1252\pm 0.0060$ & $0.1292\pm 0.0057$ & $0.1298\pm 0.0057$  \\

$\ln(10^{10} A_{s})$  &  $2.81\pm 0.12$ & $2.936^{+0.099}_{-0.12}$ & $2.873^{+0.093}_{-0.11}$ & $2.86\pm 0.10$  \\

$H_0 \, [\mathrm{km/s/Mpc}]$  &  $64.9^{+1.3}_{-1.8}$ & $68.43\pm 0.95$ & $67.2\pm 1.0$ & $67.54\pm 0.86$\\

$w_{0}$ &  $-0.37^{+0.24}_{-0.12}$ & $-0.842\pm 0.059$ & $-0.64\pm 0.10$ & $-0.675^{+0.059}_{-0.065}$  \\

$w_{\rm a}$ &  $< -0.90 \, (95\%)$ & $-0.63^{+0.30}_{-0.24}$ & $-1.30\pm 0.40$ & $-1.24^{+0.32}_{-0.28}$  \\
\hline 

$\Omega_{\rm m}$  &  $0.365^{+0.026}_{-0.013}$ & $0.3158\pm 0.0084$ & $0.337\pm 0.012$ & $0.3342\pm 0.0084$ \\

$\sigma_8$  &  $0.755^{+0.025}_{-0.028}$ & $0.805\pm 0.025$ & $0.788\pm 0.025$ & $0.789\pm 0.026$\\

$S_8$  &  $0.831\pm 0.025$ & $0.826\pm 0.024$ & $0.835\pm 0.025$ & $0.832\pm 0.026$\\

\hline
$\Delta \chi^2_{\rm min}$  &  $-8.18$ & $-6.98$ & $-13.01$ & $-27.68$\\

$\Delta AIC$  &  $-4.18$ & $-2.98$ & $-9.01$ & $-23.68$\\

$\ln \mathcal B_{ij}$ & $-1.92$ & $-0.54$ & $-4.35$ & $-10.86$\\

\hline \hline 
\end{tabular}}
\label{table_w0wa}
\end{table*}

\begin{table*}[htpb!]
\centering
\caption{Same as in Table \ref{table_w0wa}, but for the $w$CDM model.}
\renewcommand{\arraystretch}{1.55}
\resizebox{\textwidth}{!}{
\begin{tabular}{l||c|ccc} 
\hline
\textbf{Parameter} & \textbf{CMB$\kappa$\scalebox{1.2}{$\times$}LRG + DESI-DR2} & \textbf{+ PP} & \textbf{+ Union3} & \textbf{+ DESY5} \\
\hline \hline
$10^{2} \Omega_{\rm b} h^2$  &  $2.271\pm 0.036$ & $2.272\pm 0.038$ & $2.271\pm 0.038$ & $2.272\pm 0.038$  \\

$\Omega_{\rm c} h^2$  &  $0.1176\pm 0.0053$ & $0.1141\pm 0.0049$ & $0.1130\pm 0.0048$ & $0.1094\pm 0.0046$  \\

$\ln(10^{10} A_{s})$  &  $3.03\pm 0.10$ & $3.099\pm 0.099$ & $3.120\pm 0.099$ & $3.189\pm 0.098$  \\

$H_0 \, [\mathrm{km/s/Mpc}]$  &  $68.8\pm 1.3$ & $67.60\pm 0.99$ & $67.3\pm 1.0$ & $66.13\pm 0.89$\\

$w$ &  $-0.996\pm 0.046$ & $-0.951\pm 0.032$ & $-0.941\pm 0.035$ & $-0.899\pm 0.028$  \\
\hline 

$\Omega_{\rm m}$  &  $0.2962\pm 0.0058$ & $0.2993\pm 0.0054$ & $0.2996\pm 0.0056$ & $0.3020\pm 0.0055$ \\

$\sigma_8$  &  $0.808\pm 0.024$ & $0.806\pm 0.025$ & $0.807\pm 0.025$ & $0.804\pm 0.023$\\

$S_8$  &  $0.803\pm 0.024$ & $0.805\pm 0.024$ & $0.806\pm 0.025$ & $0.807\pm 0.023$\\
\hline
$\Delta \chi^2_{\rm min}$  &  $-0.15$ & $-3.96$ & $-2.83$ & $-12.15$\\

$\Delta AIC$  &  $\hspace{0.25cm} 1.85$ & $-1.96$ & $-0.83$ & $-10.15$\\

$\ln \mathcal B_{ij}$ & $\hspace{0.25cm} 2.31$ & $\hspace{0.25cm} 1.38$ & $ \hspace{0.25cm} 1.02$ & $-3.17$\\
\hline \hline 
\end{tabular}}
\label{table_DEE}
\end{table*}

\begin{figure*}[htpb!]
    \centering
    \includegraphics[width=0.85\textwidth]{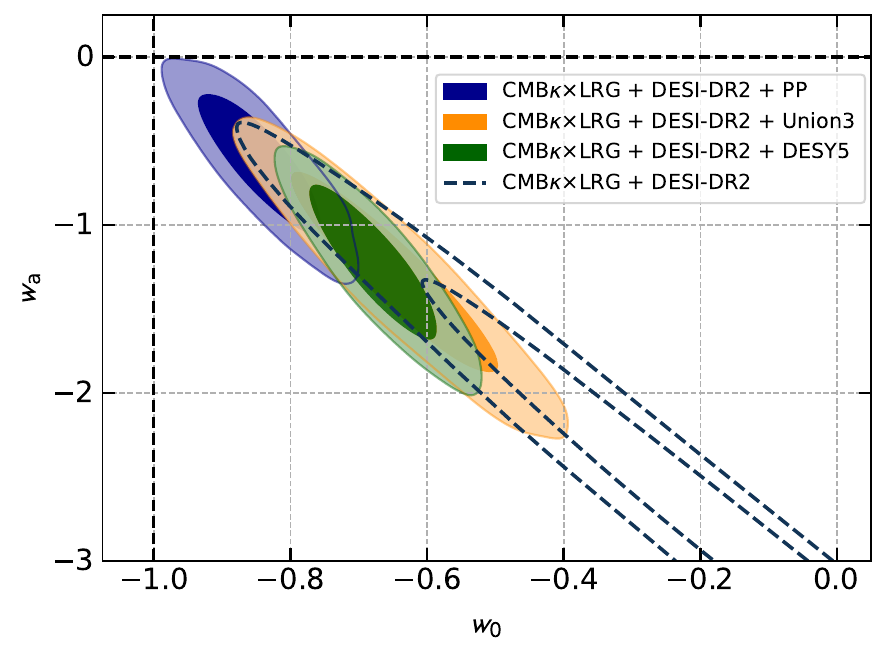}
    \caption{Two-dimensional confidence contours at the 68\% and 95\% levels for the parameters $w_0$ and $w_{\rm a}$ are shown for various data combinations, as indicated in the legend, within the framework of the $w_0w_{\rm a}$CDM model.}
    \label{fig:1}
\end{figure*}

\begin{figure*}[htpb!]
    \centering
    \includegraphics[width=1.\textwidth]{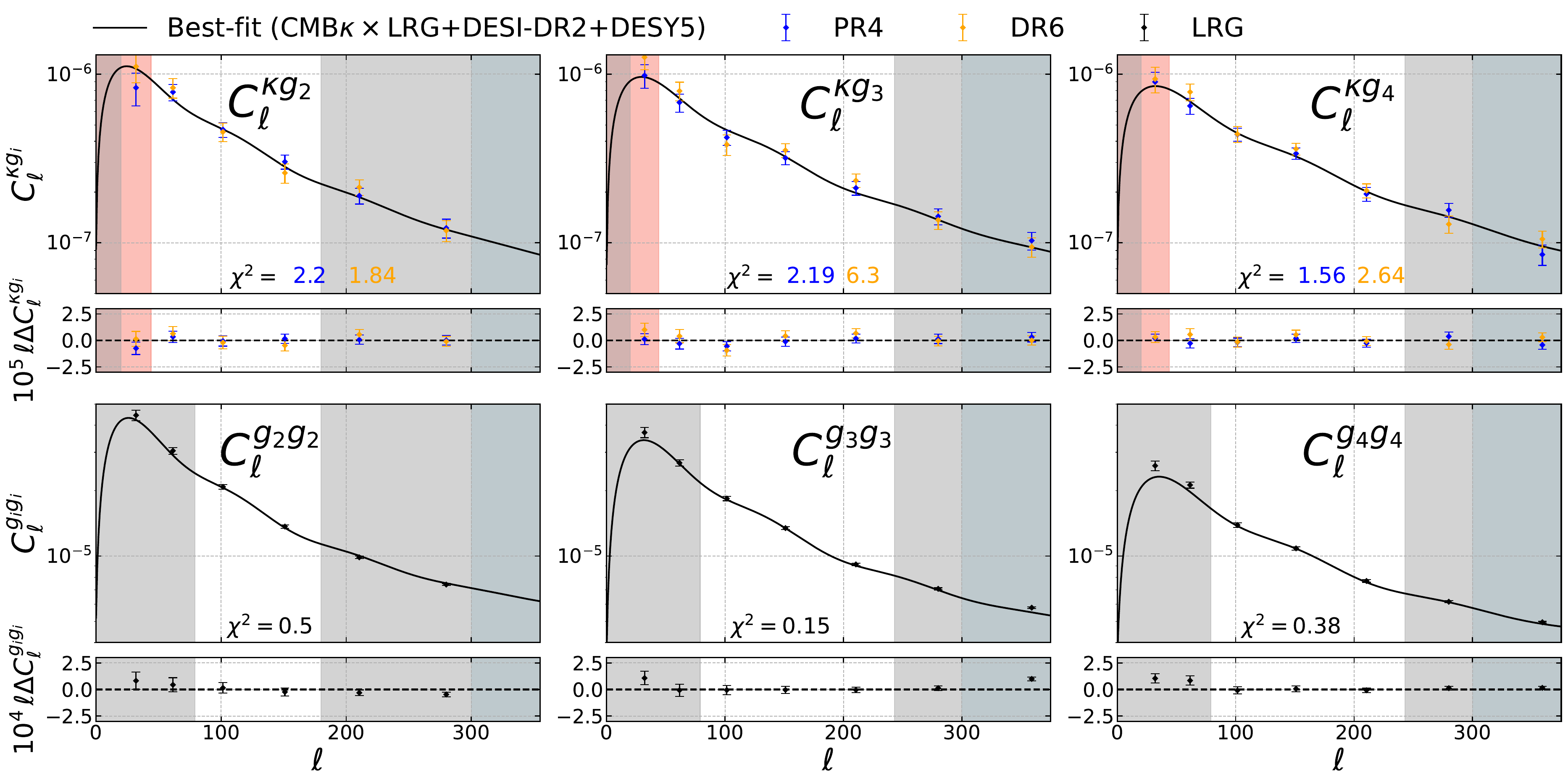}
    \caption{In the top row, we present the measurements \( C_{\ell}^{\kappa g} \) for the DESI LRG sample with Planck PR4 (blue) and ACT DR6 (orange). The third row shows \( C_{\ell}^{gg} \) for the same redshift bins. All measurements and their associated covariance matrices are taken from~\cite{Sailer:2024coh}. The solid black lines indicate the best-fit predictions obtained from our joint fit CMB$\kappa$\scalebox{1.2}{$\times$}LRG+DESI-DR2+DESY5 for $w_{0}w_{a}$CDM, using all three redshift bins within the linear theory framework described in Section~\ref{modeling}. The second and fourth rows display the residuals concerning the best-fit model for the cross- and auto-correlations, respectively. {For each measurement, we display its individual $\chi^2$ value indicated by colors.} The grey shaded regions indicate the scales excluded from our fits, while the red shaded areas represent the scales additionally excluded for ACT DR6 data, as summarized in Table~\ref{bins}. {Lastly, the blue region indicates from which $\ell_{\rm max}$ we truncate the sum when convolving our theoretical predictions with the corresponding window functions, as discussed in~\textsection\ref{cross-samples}.}}
    \label{bestfits}
\end{figure*}

\begin{figure*}[htpb!]
    \centering
    \includegraphics[width=0.85\textwidth]{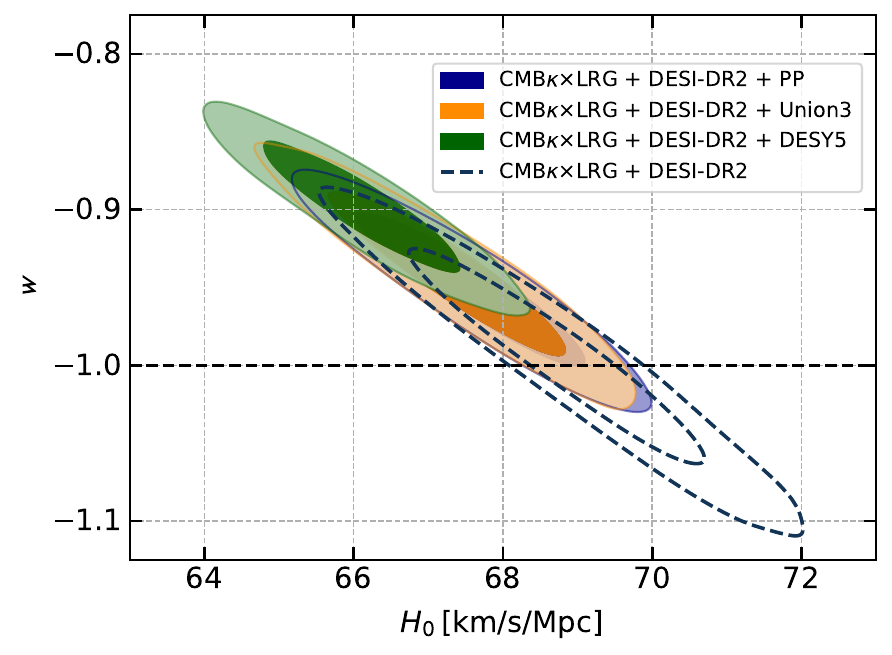}
    \caption{Two-dimensional confidence contours at the 68\% and 95\% levels for the parameters $w_0$ and $H_0$ are shown for various data combinations, as indicated in the legend, within the framework of the $w$CDM model.}
    \label{fig:2}
\end{figure*}

For the joint analysis \( \mathrm{CMB}_{\kappa} \times \mathrm{LRG} \) + DESI-DR2, we obtain \( w_0 = -0.37^{+0.24}_{-0.12} \) and a one-sided bound \( w_a < -0.90 \) $(95\%)$. Based on the statistical metrics adopted, particularly the absolute Bayesian evidence \( |\ln \mathcal{B}_{ij}| \) and the AIC, we find moderate evidence in favor of this model when compared to the standard \( \Lambda \)CDM scenario. The parameter constraints also indicate slightly higher values of \( \Omega_m \) and lower values of \( \sigma_8 \), resulting in a combined constraint of \( S_8 = 0.831 \pm 0.025 \) at 68\% CL. These results show that this joint analysis is capable of reasonably constraining the full parameter space of the model.

In what follows, we perform a series of joint analyses combining \( \mathrm{CMB}_{\kappa} \times \mathrm{LRG} \) + DESI-DR2 data with different SNIa samples, namely PP, Union3, and DESY5. In all cases, we find that the full parameter space of the \( w_0w_{\rm a} \)CDM model is well constrained.

For the combination \( \mathrm{CMB}_{\kappa} \times \mathrm{LRG} \) + DESI-DR2 + PP, we observe a mild deviation from the standard \( \Lambda \)CDM model at approximately the 95\% CL across the parameter space. However, the Bayesian evidence yields \( |\ln \mathcal{B}_{ij}| = -0.54 \), which is classified as weak evidence in favor of the \( w_0w_a \)CDM model. The joint analyses that include the Union3 and DESY5 SNIa samples lead to significant improvements in the global parameter constraints. In these cases, we find \( |\ln \mathcal{B}_{ij}| = -4.35 \) and \( -10.86 \), respectively, which correspond to very strong Bayesian evidence in favor of the dynamical dark energy scenario. In particular, the combination \( \mathrm{CMB}_{\kappa} \times \mathrm{LRG} \) + DESI-DR2 + DESY5 yields parameter estimates of \( w_0 = -0.675^{+0.059}_{-0.065} \) and \( w_a = -1.24^{+0.32}_{-0.28} \) at 68\% CL, suggesting a deviation from \( \Lambda \)CDM at the level of approximately 4.8\( \sigma \).

Figure~\ref{fig:1} presents the two-dimensional marginalized contours at 68\% and 95\% CL in the \( w_0 \)--\( w_a \) plane for the various joint analyses discussed in this work. It is particularly insightful to compare our results with the analysis by the DESI Collaboration~\cite{DESI:2025zgx}, who reported constraints from CMB + DESI DR2 + DESY5 yielding \( w_0 = -0.752 \pm 0.057 \) and \( w_a = -0.86^{+0.23}_{-0.20} \). From a statistical perspective, our constraints are fully compatible with theirs. Furthermore, the correlation trends in the \( w_0 \)--\( w_a \) plane observed in our results are consistent with those found in the DESI team's analyses, confirming that \( \mathrm{CMB}_{\kappa} \times \mathrm{LRG} \), when combined with high-quality geometric probes such as BAO and SNIa samples, provides a robust and complementary observational test for probing the nature of DE. 

Figure~\ref{bestfits} shows the success of our framework in predicting the observed \( C_{\ell}^{\kappa g} \) and  \( C_{\ell}^{gg} \) for the DESI LRG sample joint with geometric measurements.  The agreement between model and data is generally good across all redshift bins and datasets within the scales used in our fit. The residual panels (second and fourth rows) help visualize any systematic deviations between the data and the model prediction. Important to note, the residuals remain consistent with zero across most scales, suggesting that our modeling approach provides an adequate description of the data at the scales considered.

The observational nature of the \( \mathrm{CMB}_{\kappa} \times \mathrm{LRG} \) cross-correlation is particularly valuable for constraining the parameter \( S_8 \), which is one of the primary motivations for employing this cosmological probe. In the context of the \( w_0w_{\rm a} \)CDM model, we obtain \( S_8 = 0.826 \pm 0.025 \), \( 0.835 \pm 0.025 \), and \( 0.832 \pm 0.025 \) from the joint analyses with PP, Union3, and DESY5, respectively. These values are in good agreement with recent cosmic shear measurements obtained within the standard \( \Lambda \)CDM framework~\cite{Wright:2025xka,Stolzner:2025htz}. {We also note consistency with the DES shear analysis that explicitly accounts for intrinsic alignment (IA) mitigation effects \cite{DES:2024xvm}. However, our result remains in tension with the HSC Year 3 results~\cite{Sugiyama:2023fzm}, which prefer lower values of $S_{8}$.}
\\

{Recent studies have also explored constraints on DDE models independent of the primary CMB anisotropy data. A consistent trend appears across these analyses, including ours: evidence pointing to a phantom-like behavior of DE at early times, transitioning to a quintessence-like regime at late times around $z\sim0.33$. More precisely, Ref.~\cite{Maus:2025rvz} yield $w_0 = -0.755^{+0.083}_{-0.055}$ and $w_a = -1.03^{+0.25}_{-0.36}$ using $P_{\ell}(k)+\xi_{\ell}^{\text {post }}(s)+C_{\ell}^{g \kappa} +$ Union3, in agreement with our results from the analysis using Union3. Additionally, Ref.~\cite{Silva:2025twg} reports $w_0 = -0.692 \pm 0.064$ and $w_a = -1.18 \pm 0.30$, using the Full-Shape (FS) galaxy power spectrum from the BOSS DR12 sample in conjunction with DESI-DR2 + DESY5, which agrees with the one in this work. Lastly, our result using the PP dataset is also in agreement with the values report in Ref.~\cite{Chen:2025jnr}, where CMB + DESI-DR2 + PP is combined with the FS galaxy power spectrum and bispectrum data from BOSS.}


\subsection{$w$CDM}
Figure~\ref{fig:2} presents the two-dimensional marginalized contours at 68\% and 95\% CL in the $w-H_0$ plane for the various joint analyses for the $w$CDM model, that is, assuming $w_a = 0$. {Therefore, the equation of state (EoS) remains constant throughout the entire cosmic evolution}. The statistical constraints for the parameters are shown in the table~\ref{table_DEE}. 

For all analyses performed within the \( w \)CDM framework, the statistical evidence for deviations from the standard \( \Lambda \)CDM cosmology is considerably reduced. We find that only the joint analysis \( \mathrm{CMB}_{\kappa} \times \mathrm{LRG} \) + DESI-DR2 + DESY5 provides moderate-level evidence for a departure from the cosmological constant scenario, yielding \( w = -0.899 \pm 0.028 \). {This result points to a clear preference for a quintessential dark energy component (\( w > -1 \)). It is important to note that, in this specific case, the EoS remains physically unchanged throughout cosmic expansion.} The other joint analyses incorporating PP and Union3 SNIa samples also favor for a (\( w > -1 \)), though with only weak to positive evidence. {The only joint analysis statistically compatible with \( w = -1 \) is \( \mathrm{CMB}_{\kappa} \times \mathrm{LRG} + \) DESI-DR2. This whole discussion becomes clear upon examining Figure~\ref{fig:2}.} Regarding the Hubble constant \( (H_0) \), all our analyses consistently yield values around \( H_0 \simeq 67.60 \,\mathrm{km\,s^{-1}\,Mpc^{-1}} \), which are in excellent agreement with estimates from early-time CMB anisotropy data. Therefore, we conclude that the \( w \)CDM model, even when constrained by \( \mathrm{CMB}_{\kappa} \times \mathrm{LRG} \), does not alleviate the current Hubble tension.

\section{Final Remarks}
\label{conclusion}

In this work, we have investigated the cosmological implications of the \( \mathrm{CMB}_{\kappa} \times \mathrm{LRG} \) cross-correlation in combination with complementary observational data sets, focusing on constraints within the \( w_0w_{\rm a} \)CDM and \( w \)CDM frameworks. We have shown that combining \( \mathrm{CMB}_{\kappa} \times \mathrm{LRG} \) with recent geometric measurements can lead to strong statistical evidence in favor of a dynamical dark energy scenario. Our results demonstrate that the inclusion of this cross-correlation observable plays a significant role in improving parameter constraints, especially on the clustering amplitude parameter \( S_8 \), and offers valuable insights into the nature of DE. 

We find that while the \( w \)CDM model yields results broadly consistent with the \( \Lambda \)CDM scenario, only the most comprehensive joint analysis, \(\mathrm{CMB}_{\kappa} \times \mathrm{LRG}\)+DESI-DR2+DESY5, provides moderate statistical evidence for a deviation from a cosmological constant, with a preference for a quintessential DE component. However, our findings indicate that the \( w \)CDM model remains unable to address the current Hubble tension, as all estimates for \( H_0 \) remain consistent with those inferred from early-universe data.

Overall, our analysis highlights the importance of the \( \mathrm{CMB}_{\kappa} \times \mathrm{LRG} \) cross-correlation as a powerful observational probe. We have shown that it can be effectively combined with geometric measurements to test more complex and physically motivated cosmological models. As upcoming surveys such as the LSST \cite{LSSTScience:2009jmu}, Euclid \cite{Euclid:2019clj}, and CMB-S4 \cite{CMB-S4:2016ple} come online, the precision and coverage of both lensing and galaxy clustering measurements will dramatically improve. These future datasets will enable even more stringent tests of dynamical dark energy, modified gravity, and other extensions to the standard cosmological model, with \( \mathrm{CMB}_{\kappa} \times \mathrm{LRG} \) cross-correlations poised to play a central role in this endeavor.
\\

\textbf{Data Availability}: The datasets and products underlying this research, such as Boltzmann codes and likelihoods, will be available upon reasonable request to the corresponding author after the publication of this article.
\\

\begin{acknowledgments}
\noindent {We thank the referee for the thoughtful feedback, which helped enhance both the clarity and the impact of our results.} We are also grateful to Martin White and Noah Sailer for valuable discussions regarding the galaxy–CMB lensing cross-correlation samples used in this work. M.A.S received support from the CAPES scholarship. R.C.N. thanks the financial support from the Conselho Nacional de Desenvolvimento Científico e Tecnologico (CNPq, National Council for Scientific and Technological Development) under the project No. 304306/2022-3, and the Fundação de Amparo à Pesquisa do Estado do RS (FAPERGS, Research Support Foundation of the State of RS) for partial financial support under the project No. 23/2551-0000848-3.
\end{acknowledgments}

\appendix
\section{Supplementary Pipeline Verifications} 
\label{Appendix}

{To ensure that our main results are robust against modeling assumptions, we performed a series of internal consistency checks and pipeline variations. This appendix summarizes these tests and quantifies their impact on the cosmological constraints relative to our baseline comparison analysis presented in~\textsection\ref{cross-samples}.}

{To quantify the level of agreement or tension between the analyses, we adopt the quadratic estimator introduced in~\cite{Addison:2015wyg}, defined as follows:}
\begin{equation}
    \chi^2 = \left(\mathbf{x}_i - \mathbf{x}_j\right)^\mathrm{T} \left(\mathcal{C}_i + \mathcal{C}_j\right)^{-1} \left(\mathbf{x}_i - \mathbf{x}_j\right),
    \label{N-tension}
\end{equation}
{where $ \mathbf{x}_i $ and $ \mathbf{x}_j $ are vectors containing the mean values of the cosmological parameters inferred from the analyses $ i $ and $ j $, respectively. The quantities $ \mathcal{C}_i $ and $ \mathcal{C}_j $ denote the corresponding covariance matrices associated with the parameter estimates for each analysis. This formalism provides a statistically rigorous measure of consistency between two analyses, reflecting how well they align within the parameter space of the theoretical model. By comparing analyses in terms of their mean values and associated uncertainties, the method helps to identify discrepancies that may point to systematic errors, inconsistencies, or the need for modifications to the underlying theoretical framework.}

\begin{figure*}[htpb!]
    \centering
    \includegraphics[width=0.75\textwidth]{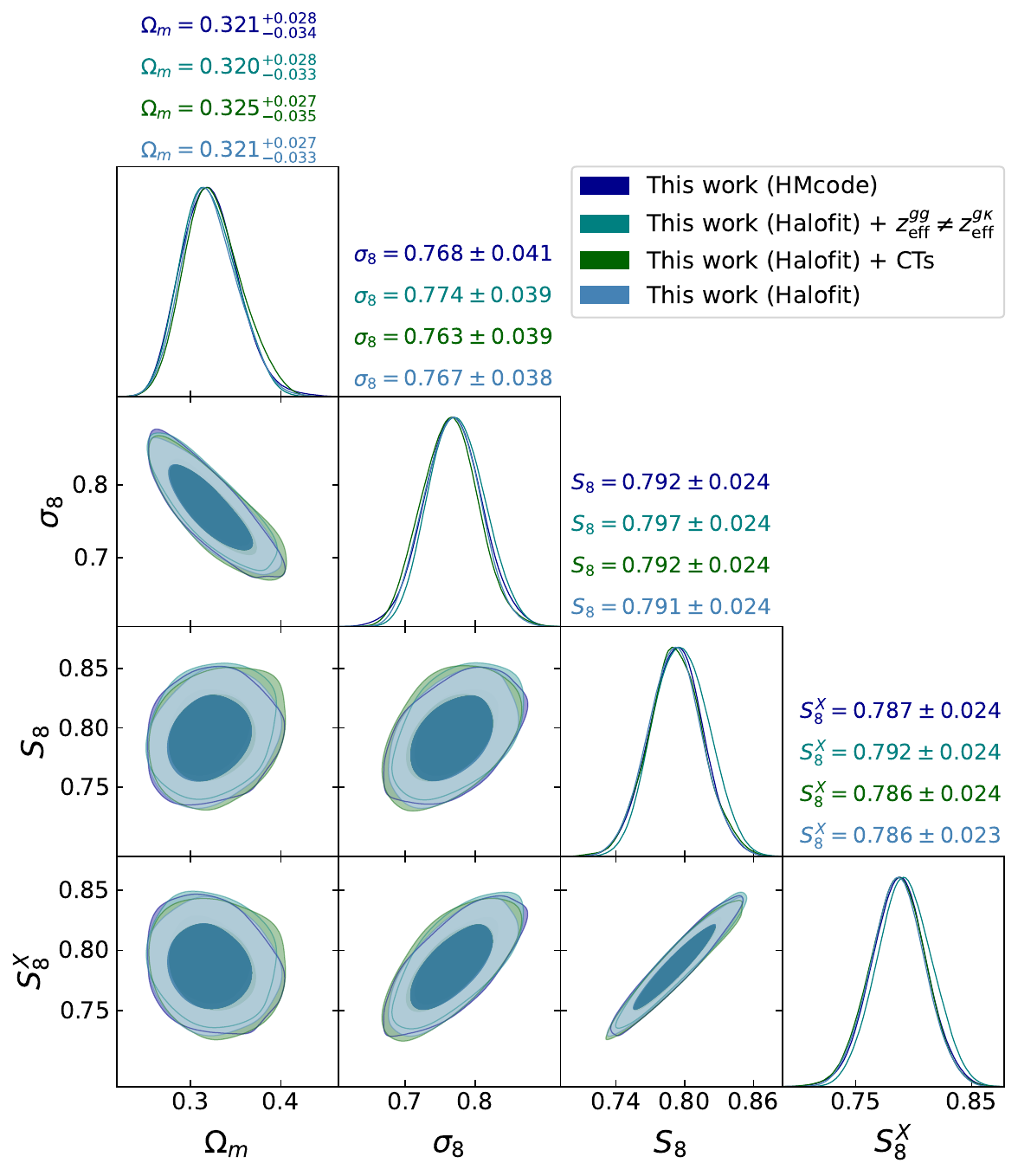}
    \caption{{Triangle plot comparing the posterior distributions of cosmological parameters obtained in this work using our baseline approach explained in~\textsection\ref{cross-samples} with some pipeline variations. The contours correspond to the 68\% and 95\% confidence levels. The agreement between all the results serves as a consistency check and confirms the accuracy of our likelihood implementation.} }
    \label{linear_theory_extended}
\end{figure*}

{For all tests shown in Figure~\ref{linear_theory_extended}, we perform the MCMC analysis using the same priors and likelihood structure described in~\textsection\ref{data}, with the appropriate variations. The resulting shifts in cosmological parameters are statistically negligible, confirming the robustness of our main conclusions. We summarize the main variants below:}

\begin{itemize}
    \item \textbf{HMcode vs Halofit}: We replaced the nonlinear matter power spectrum from Halofit to HMcode (as implemented in CLASS),  which includes a physically motivated treatment of baryonic effects and BAO damping. We find that this change only introduces a negligible variation of $0.08\sigma$ in our analysis.

    \item \textbf{Adding counterterms (CTs)}: Following the modeling in \cite{Sailer:2024coh}, we included the counterterm contributions proportional to $k^2 P(k)$ in the theoretical predictions of the angular spectra, with the priors $\alpha_{\rm a}(z_{i}) \in \mathcal{N}(0,3)$ and $\alpha_{\rm x}(z_{i}) \in \mathcal{N}(0,3)$. Again, we find that this change only introduces a small shift of $\lesssim 0.1\sigma$ in our analysis.
    
    \item {Different effective redshifts}: In our baseline, we used the same effective redshift per bin to evaluate both observables. In this variant, we used kernel-weighted effective redshifts specific to each observable \cite{Chen:2022jzq}, meaning Eq.~\ref{effective_z} for $C^{gg}$ and Eq.~$2.6$ from \cite{Modi:2017wds} for $C^{g\kappa}$. This led to a mid $0.18\sigma$ variation.
\end{itemize}

{In general, these changes can be interpreted as minor statistical fluctuations of purely numerical origin, rather than indicative of any physical effect. Such variations arise naturally from the use of different numerical implementations and sampling algorithms inherent to MCMC analyses.}

\bibliographystyle{apsrev4-1}
\bibliography{bib}

\end{document}